\documentclass{emulateapj}
\usepackage{apjfonts}
\usepackage{times}
\usepackage{graphicx}
\usepackage{color}
\begin{document}
\title{Correlation between Global Parameters of Galaxies}
\author{Yu-Yen Chang \altaffilmark{1,2,3}, Rikon Chao \altaffilmark{2,4}, Wei-Hao Wang \altaffilmark{3}, Pisin Chen \altaffilmark{1,2,5}}
\affil{1. Department of Physics and Graduate Institute of Astrophysics, National Taiwan University, Taipei, Taiwan 10617}
\affil{2. Leung Center for Cosmology and Particle Astrophysics, National Taiwan University, Taipei, Taiwan 10617}
\affil{3. Institute of Astronomy and Astrophysics, Academia Sinica, Taipei, Taiwan 10617}
\affil{4. Department of Electrical Engineering, National Taiwan University, Taipei, Taiwan 10617}
\affil{5. Kavli Institute for Particle Astrophysics and Cosmology, SLAC National Accelerator Laboratory, Stanford University, Stanford, CA 94305, U.S.A.}
\begin{abstract}
Recently Disney et al. (2008) found a striking correlation among the five basic parameters that govern the galactic dynamics: $R_{50}$, $R_{90}$, $L_r$, $M_d$, and $M_{HI}$. They suggested that this is in conflict with the $\Lambda$CDM model, which predicts the hierarchical formation of cosmic structures from bottom up. Considering the importance of the issue, we performed a similar analysis with a significantly larger database and two additional parameters, $L_J$ and $R_J$, of the near-infrared $J$ band. We used databases from the Arecibo Legacy Fast Arecibo $L$-band Feed Array Survey for the atomic gas properties, the Sloan Digital Sky Survey for the optical properties, and the Two Micron All Sky Survey for the near-infrared properties, of the galaxies. We found that the first principal component dominates the correlations among the five parameters and can explain 83\% of the variation in the data. When color ($g-i$) is included, the first component still dominates and the color forms a second PC that is almost independent of other parameters. The overall trend in our near-infrared PCA is very similar, except that color ($i-J$) seems even more decoupled from all other parameters. The dominance of the first PC suggests that the structure of galaxies is governed by a single physical parameter. This confirms the results in Disney et al. However, based on the importance of the baryon physics in galaxy formation, we find it premature to conclude that the hierarchical structure formation scenario and the notion of cold dark matter are necessarily flawed.
\end{abstract}
\section{Introduction}
One way to understand our universe is to gain insights into the structure of galaxies. For one thing, it helps to reveal the role of dark matter in their formation and dynamics. The cosmological model that consists of a cosmological constant and the cold dark matter in addition to the ordinary baryon matter and radiation ($\Lambda$CDM) has been able to successfully explain the evolution of the cosmic structures especially at large scales. By measuring CMB fluctuations \citep[COBE;][]{1992ApJ...396L...1S,1994ApJ...436..423B} \citep[WMAP;][]{2003ApJS..148....1B,2003ApJS..148...97B,2009ApJS..180..330K,2009ApJ...701.1804D}, Type Ia supernovae \citep{1998AJ....116.1009R} and gravitational lensing \citep{1937ApJ....86..217Z}, this model of cosmology has by now been established as the standard model of cosmology. Among its various implications, it suggests a hierarchical, bottom-up history of structure formation that evolves from small fluctuations to galaxies, clusters and eventually superclusters. 

At small scales, the success of $\Lambda$CDM has not been as clearcut. There still exist several inconsistencies between $\Lambda$CDM and the observations at small scales. For example, the simulations based on $\Lambda$CDM have revealed more number of galactic satellites \citep{1999ApJ...524L..19M,1999ApJ...522...82K,2009NJPh...11j5029P,2010AdAst2010E...8K} and less number of disk galaxies \citep{2001ApJ...551..608S} than what have been observed. Besides, the degree of emptiness in the voids is also inconsistent between theory and observation \citep{2001ApJ...557..495P}. While the $\Lambda$CDM can explain the galaxy rotational curves at large radii \citep{1997MNRAS.290..533D,2009arXiv0912.1668F}, the relatively higher density at the galactic core than that predicted by the $\Lambda$CDM, that is, the so-called cusp-core problem, is still unresolved \citep{2001ApJ...561...35D}. Aside from these, there is the newly emerged challenge to the CDM model with regard to the hierarchical galactic structure formation. That is, if galaxies are the results of random hierarchical merger processes, their basic parameters are expected to be independent. Yet a recent study found that the basic parameters of galaxies are highly correlated.

The hierarchical galaxy formation scenario has been actively investigated previously \citep[e.g.,][]{1991ApJ...379...52W, 2000MNRAS.319..168C, 2006RPPh...69.3101B}, and the correlation between selected galactic parameters has been studied in the past \citep{1976ApJ...204..668F,1977A&A....54..661T,1996A&A...312..397G,2008ApJ...682..861B}. However, the overall correlation between all major galactic variables has never been investigated until recently by (\citealt{2008Natur.455.1082D}; hereafter D08), who made the remarkable finding, based on 195 galaxies, that there exist very strong correlations between six galactic variables: the 90\%-light radius ($R_{90}$), the 50\%-light radius ($R_{50}$), the \ion{H}{1} mass ($M_{HI}$), the dynamical mass ($M_d$), the luminosity ($L$), and the color. They suggested that this was in conflict with the notion of the hierarchical structure formation scenario, which would predict more collisions among substructures and therefore less regularities between galaxies. Since CDM model is the underlying tenet of the hierarchical scenario, the authors assert that CDM model, or even the notion of dark matter itself, is in doubt.

Considering the importance of this issue, we set out to reinvestigate the analysis made by D08. We note that their claim actually involves two separate sub-issues. First, there is the issue of whether the galactic dynamics is indeed controlled by a single parameter. Second, even if this is true, there still remains the issue of whether such a fact necessarily concludes the failure of the hierarchical structure formation scenario and the cold dark matter model. In our view the positive conclusion of the former does not necessarily lead to the affirmative conclusion of the latter. In this paper we demonstrate tentative evidence that there may exist more than one principal component among the global parameters of galaxies with regard to the first issue. As for the second issue, we argue about the importance of the non-gravitational baryon physics in galactic structure formation, which renders the naive extrapolation of the hierarchical structure formation scenario from cosmic to galactic scales questionable. 

In order to scrutinize the correction issue, we performed a similar analysis on global parameters of galaxies with a significantly larger database and two additional parameters, $L_J$ and $M_J$, based on the infrared $J$ band. We include a total of 1022 galaxies from the Arecibo Legacy Fast Arecibo $L$-band Feed Array Survey \citep[ALFALFA;][]{1996PASA...13..243S,2007AJ....133.2569G,2008AJ....135..588S,2008AJ....136..713K} for the atomic gas properties, and the Sloan Digital Sky Survey \citep[SDSS;][]{2009ApJS..182..543A,2009yCat.2294....0A} for the optical properties. Among the 1022 galaxies, 481 of them have also been detected by the Two Micron All Sky Survey \citep[2MASS;][]{2006AJ....131.1163S}. We use these galaxies to study their near-infrared properties.

Recently, several large blind 21cm surveys for galaxies have been conducted. Extragalactic \ion{H}{1} surveys provide a new way of investigating galaxy properties other than the stellar luminosity, surface brightness and morphology. Because of the largeness of the database, the correlations between \ion{H}{1} and other properties can be studied \citep{1984AJ.....89..758H,2005AJ....129.1311R}. The \ion{H}{1} observation has by now been established as a useful tracer to study galaxies, especially those with low surface brightness. In addition, the dynamics of the neutral hydrogen gas in principle constrains the dark matter content of the galaxies. It therefore makes sense to include the \ion{H}{1} data in the galactic parameter correlation studies with the hope that it would provide us additional information about galaxy formation and evolution.

Near-infrared studies of \ion{H}{1} selected galaxies had been attempted by various groups \citep[e.g.,][]{2005AJ....129.1311R,2009MNRAS.394..340G}, but with relatively small samples.  Our motivation of adding the near-infrared data in our analysis is the following. The optical emission is sensitive to young stars. The near-infrared emission, on the other hand, is less affected by the young stars and is therefore a better tracer of the total stellar mass, which dominates the baryonic matter at galactic scales. We believe that the inclusion of the infrared data would provide us additional and independent information on the baryonic mass assembly history of the galaxies. 

Through the principal component analysis \citep[PCA;][]{jolliffe}, we confirm that except the color, all other observables, from \ion{H}{1}, optical to near-infrared bands, are highly correlated and dominated by a single parameter. This is true both in the optical and in the near-infrared bands and this confirms the results in D08. However, based on the importance of the non-gravitational baryon physics in galaxy formation, we are not convinced that the hierarchical structure formation scenario and the notion of cold dark matter are necessarily flawed.

The organization of this paper is as follows. We describe the data and sources in \textsection 2. Then several variables are adopted and apply to statistical analysis in \textsection 3. In \textsection 4, we summarize and discuss our results. 
\section{Data}
Our samples are the blind 21 cm survey from ALFALFA. This selects gas rich galaxies which also contain low luminosity and low surface brightness galaxies in higher proportion than those in an optical selection. The optical data for this study come from the SDSS DR7, which covers 12,000 deg$^2$ for imaging and provides spectra of 930,000 galaxies. Here we briefly describe how we select SDSS counterparts to the ALFALFA sources, and we refer to \citet{2009MNRAS.394..340G} for more detailed discussion about identification. 

The Arecibo Telescope has a beam size of $3\farcm5$ at 21 cm. Since the majority of the \ion{H}{1} detections have S/N $>10$ \citep{2007AJ....133.2569G,2008AJ....135..588S,2008AJ....136..713K}, we adopted a conservative searching radius of $10\arcsec$. We found 1233 SDSS galaxies that appear to be detected by ALFALFA. We then excluded Virgo galaxies, because their neutral hydrogen is known to undergo strong environmental impact (e.g., \citealp{2009AJ.138.1741C}). We also excluded galaxies whose half-light radii are too small ($<1\arcsec$) comparing to the SDSS resolution. Such small half-light radii either arise from mis-identifications (from stars) or would result in large uncertainties.  We are left with a large sample of 1022 SDSS galaxies.  Among these galaxies, 889 have $g$ magnitudes that are $<18$ and 120 have $g=18$--20. The cumulative number counts of SDSS galaxies \citep[e.g.,][]{2001AJ....122.1104Y} are $\sim60$ deg$^{-2}$ at $g<18$ and $\sim450$ deg$^{-2}$ at $g=18$--20. Given our $10\arcsec$ search radius, we therefore expect at most 1.3 mis-identifications in our 889 $g<18$ galaxies, and additional 1.3 mis-identifications in our 120 $g=18$--20 galaxies.  These number are sufficiently small and mis-identified galaxies should not impact our analyses.

The total stellar masses of galaxies are more directly reflected by the near-infrared observations. We added the data from 2MASS to our samples. 2MASS provides $J$, $H$, and $K_s$-band observations of the entire sky as well as a point source catalog and an extended source catalog. We use the 2MASS All-Sky Extended Source Catalog (XSC) to find the galaxies in the \ion{H}{1} samples.  To understand the quality of the identification, we first compared the 2MASS and SDSS coordinates of the 2MASS detected galaxies.  We found that more than 90\% of them have offsets between 2MASS and SDSS that are well within their half-light radii.  We visually inspected all galaxies with large offsets of $>2\arcsec$, and found small number of cases that are likely mis-identifications as well as ongoing mergers. We excluded these galaxies from our samples.  Because 2MASS is shallower than SDSS, we are left with 481 reliably identified galaxies in the near-infrared.

From the ALFALFA released catalog 1, 2 and 3, we obtained 1796 \ion{H}{1} data, out of which 1265 galaxies could be found in the SDSS DR7 database. There are 32 galaxies within this set that are too faint in the optical to have reliable magnitudes and luminosities. Hence we finally used the remaining 1022 galaxies in our analyses. We have also analyzed the 195 galaxies of D08 from HIPASS \citep{2004MNRAS.350.1195M,2004AJ....128..502A,2009MNRAS.394..340G,2009AJ....138..796W,2010AJ....139..315W} and have obtained similar results. However, since the definitions of the observational variables, such as that of the rotational velocity, are not entirely consistent between the two data sets, we only report on the results from ALFALFA. These 1022 galaxies can be regarded as a blind \ion{H}{1} selected sample. We deduced from the data six variables, which are $R_{50}$ (half-light radius in units of pc) $R_{90}$ ($90\%$-light radius in units of pc), $L_r$ (luminosity in $r$ band in solar units), $M_{HI}$ (\ion{H}{1} mass in solar units), $M_d$ (dynamical mass in solar units) and color ($g-i$).

The variables $R_{50}$ and $R_{90}$ represent the radii in the Petrosian system \citep{1976ApJ...209L...1P,2001AJ....121.2358B, 2001AJ....122.1104Y}. In SDSS, the parameters are {\tt petroR50} and {\tt petroR90}, respectively. Because the Petrosian system is based on circular objects, we further corrected the radii with the major-to-minor axis ratios, which are the parameters {\tt deVAB\_r} or {\tt expAB\_r} in SDSS. To do this, we follow the result in \citet{westth} and \citet{2009AJ....138..796W,2010AJ....139..315W}. The authors fitted the corrections from circular to elliptical apertures as functions of major-to-minor axis ratios. We directly adopted their formulas for our corrections. By comparing the likelihoods of the de Vaucouleur and the exponential models, we chose the one with the larger likelihood between {\tt deVAB\_r} and {\tt expAB\_r}. $L_r$ was derived from the Petrosian system and calculated from the Petrosian magnitude, {\tt petroMag\_r}. Color is from the model magnitude, {\tt modelMag\_g} and {\tt modelMag\_i}, which is suggested by the SDSS database. The variable $M_d$ is calculated by $(\Delta V)^2 R_{90}/G$, in which $\Delta V$ is the rotational velocity from the \ion{H}{1} spectra and corrected for the inclination with the major-to-minor axis ratio as what we did for $R_{50}$ and $R_{90}$. $M_{HI}$ is acquired directly from the ALFALFA database and it is derived from the \ion{H}{1} flux. 

Because there is high degeneracy between $J$, $H$, and $K_s$ bands, we only chose $J$ band to represent the near-infrared data. Therefore, we acquired $R_J$ (half-light radius in $J$ band in units of pc) and $L_J$ (luminosity in $J$ band in solar units) of the 481 galaxies that overlap in the ALFALFA, SDSS and 2MASS catalogs to gain a better insight into the stars of the galaxies. More specifically, $L_J$ is derived from the magnitude in $J$ band, which is the parameter {\tt j\_m\_ext} in the 2MASS database. This magnitude is based on the extrapolation of the fit to the surface brightness profile. And $R_J$ is the integrated half-flux radius of $J$ band, which is the parameter {\tt j\_r\_eff} in the 2MASS database.
\section{Methods and Results}
\begin{figure}[ht]
\begin{center}\includegraphics[width=1\columnwidth]{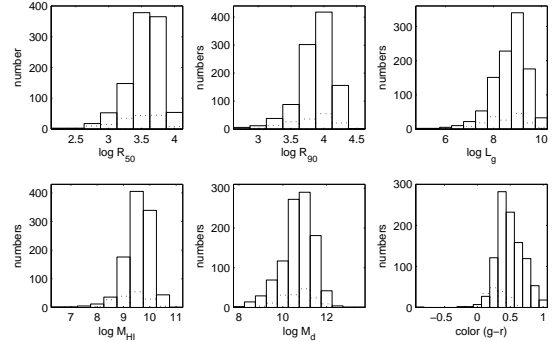}\end{center}
\caption{Distribution of the variables. The solid histograms are our samples and the dashed histograms are the samples of D08. We could only find 157 galaxies with rotational velocities for $M_d$ in D08 and \citet{2009MNRAS.394..340G}.}
\label{hist}
\end{figure}

Our sample of 1022 galaxies is not only larger than that in D08 and \citet{2009MNRAS.394..340G}, but also covers broader ranges of size, luminosity, and mass (Figure \ref{hist}). The minimum value of $L_r$ in our sample is much smaller than that of D08. As for the \ion{H}{1} and the dynamical masses, although the median values of the two samples are similar, our sample contains a substantial amount of lower mass galaxies. Our sample also covers a broader range of the $g-r$ color. Because of the larger sampling and the wider range of the galactic properties, our sample is in general more representative. However since 2MASS is shallower, our 2MASS subsample of 481 galaxies is relatively speaking less representative than that from SDSS. Even so, our 2MASS subsample is still substantially larger than that of D08.

It is important to investigate whether our \ion{H}{1} selected sample is biased against early type galaxies, since such galaxies are usually gas poor. To do so, we identify spheroidal and disk galaxies in our sample based on the morphology with a method similar to that in \citet{2001AJ....122.1861S}. In the SDSS database, there are de Vaucouleur and exponential models for each galaxy. By comparing the likelihood and the fractions of the two models for our 1022 galaxies, we found that 804 galaxies are disk-like and 218 galaxies are spheroidal-like. 

\begin{figure}[ht]
\begin{center}\includegraphics[width=1\columnwidth]{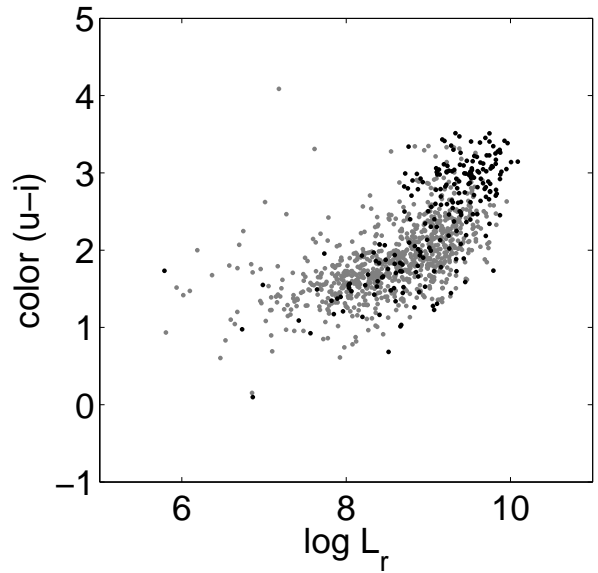}\end{center}
\caption{The $u-i$ vs.\ $L_r$ diagram. Gray dots are disk galaxies that are better described by exponential models. Black dots are spheroidal galaxies that are better described by de Vaucouleur models.}
\label{fig_sph_disk}
\end{figure}

In Figure \ref{fig_sph_disk}, we show a color--luminosity diagram for our 1022 galaxies. The spheroidal galaxies are in general more luminous, and redder than the disk galaxies. This is consistent with what we expect for elliptical and spiral galaxies. Most importantly, in the color-luminosity space, the spheroidal galaxies are redder than the blue cloud although they do not yet form a complete red sequence. Our sample thus appears to include a fair number of red and elliptical galaxies. Although the bias against extremely gas-poor galaxies can be hardly avoided here, fortunately, we found no major difference between these two types in our subsequent studies. We therefore believe that the omission of extremely gas-poor galaxies should not have caused major systematic bias in our analysis.

\begin{figure}[ht]
\begin{center}\includegraphics[width=1\columnwidth]{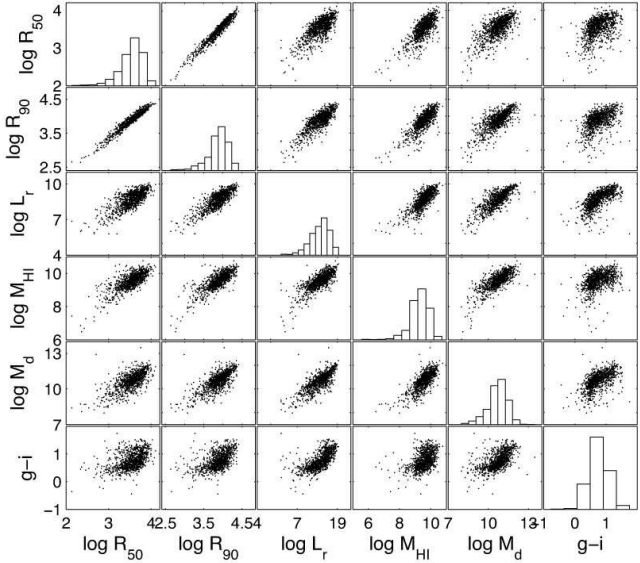}\end{center}
\caption{Scatter plots showing correlations between six measured variables. All the variables are in solar units and with logarithmic representation. The diagonal line is the histograms, which have vertical scales from 0 to 600.}
\label{fig_cor6}
\end{figure}

\begin{figure}[ht]
\begin{center}\includegraphics[width=1\columnwidth]{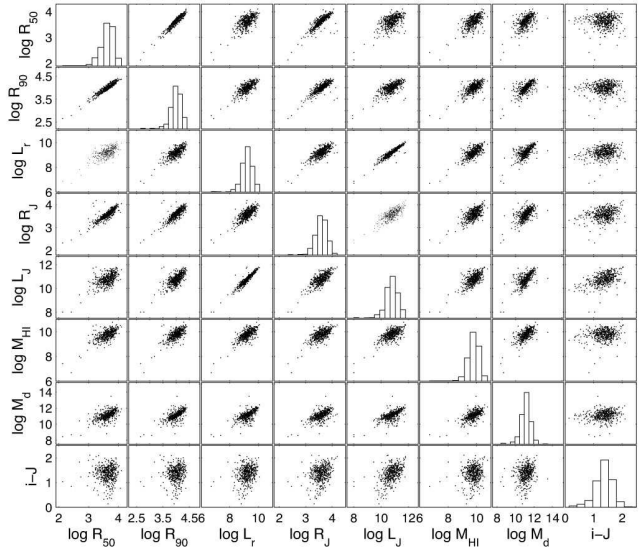}\end{center}
\caption{Scatter plots showing correlations between eight measured variables, including 2MASS data and reducing to 481 galaxies. All the variables are in solar units and with logarithmic representation. The diagonal line is the histograms, which have vertical scales from 0 to 250. There are small numbers of outliers in many of the plots. They are likely mis-identifications or bad photometry, and they do not impact our analyses.}
\label{fig_cor8}
\end{figure}

Relations among the key parameters can be inferred from the 1022 galaxies in both ALFALFA and SDSS. For instance, it is found that the half light radius is proportional to the 90\%-light radius \citep[$R_{50}\propto R_{90}$;][]{1988A&A...192..117V}; the $r$-band luminosity is proportional to the cubic power of the half-light radius \citep[$L_r\propto R_{50}^3$;][]{2003ApJ...592..819B}; the \ion{H}{1} mass is proportional to the square of the half-light radius \citep[$M_{HI}\propto R_{50}^2$;][]{1984AJ.....89..758H,2005AJ....129.1311R}, and finally, the dynamical mass is proportional to the $r$-band luminosity \citep[$M_d \propto L_r$;][]{1996A&A...312..397G}. We found that all the correlations are evident even after including the near-infrared variables, except the color (Figure \ref{fig_cor6} and Figure \ref{fig_cor8}).


As a whole, the correlations between color and other variables are much weaker than other correlations. We tested various combinations of colors and found that $g-i$ gives a larger PCA correlation coefficient than other colors (e.g., $g-r$, adopted by D08). The reason could be the larger wavelength difference between $g$ and $i$. Among the three 2MASS bands, the result based on $J$ is somewhat better in the PCA, possibly because of the better signal-to-noise ratio. Hence, we adopted $i-J$ for the color when we analyzed the 481 galaxies in 2MASS. Nevertheless, our analyses show that all the correlation coefficients are smaller than 0.7, indicating that they are not so highly correlated with other parameters. Intuitionally, more luminous galaxies tend to be redder because their colors are dominated by older stars. In fact, the color is more complex than any other variable because of the bias introduced by the very luminous young stars.

\begin{figure}[ht]
\begin{center}\includegraphics[width=1\columnwidth]{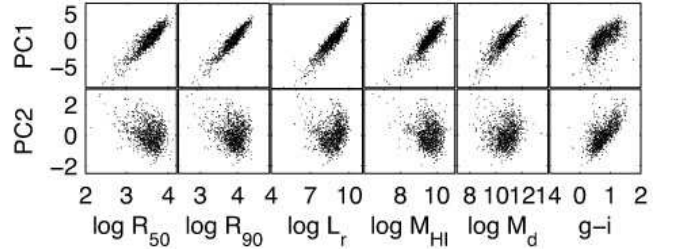}\end{center}
\caption{The PCA results for 1022 galaxies from ALFALFA and SDSS with colors. Here we only show the strongest ones, PC1 and PC2, because other principal components are not significant by PCA criterions. PC1 is well correlated with all the variables. In the first row, the color is still correlated with the other five variables and PC1. In the second row, the rightmost plot shows that the color is even more strongly correlated with a new principal component, PC2, which is not correlated with other variables. In this plot, PC1= 0.42$\log R_{50}$ + 0.44$\log R_{90}$ + 0.43$\log L_r$ + 0.42$\log M_{HI}$ + 0.41$\log M_d$ + 0.31($g-i$) and PC2= $-0.34\log R_{50}$ $-0.24\log R_{90}$ $+0.12\log L_r$ $-0.24\log M_{HI}$ + 0.08$\log M_d$ +0.86($g-i$).}
\label{pc6}
\end{figure}

\begin{figure}[ht]
\begin{center}\includegraphics[width=1\columnwidth]{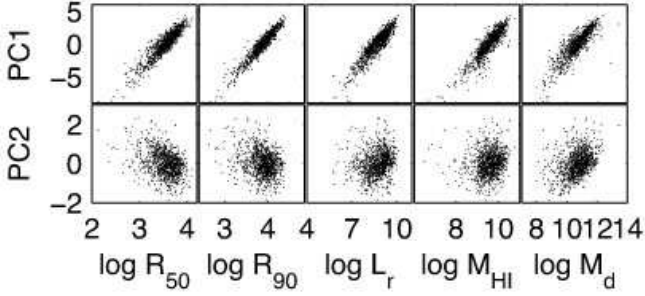}\end{center}
\caption{The PCA results for 1022 galaxies from ALFALFA and SDSS without colors. Here we only show the strongest ones, PC1 and PC2, because other principal components are not significant by PCA criterions. PC1 is well correlated with all the variables. In this plot, PC1= 0.45$\log R_{50}$ + 0.47$\log R_{90}$ + 0.45$log L_r$ + 0.44$\log M_{HI}$ + 0.43$\log M_d$ and PC2= $- 0.59\log R_{50}$ - 0.43$\log R_{90}$ + 0.28 $\log L_r$ + 0.18$\log M_{HI}$ + 0.60$\log M_d$.}
\label{pc5}
\end{figure}

We conducted PCA to find correlations among the variables. PCA typically produces a series of new variables called the principal components, namely PC1, PC2.., etc. The correlations between these principal components and the original variables then reveal the general correlations between the particular variable and the others. In our case we found that the first principal component, PC1, is highly correlated with the six observational variables. We notice that while the color is less correlated to PC1, possibly because of the recent star formation, it is much more correlated with the second principal component, PC2. In addition to the investigation into the diagram of correlations, the eigenvalues of the correlation matrices of the original variables give quantitative information of the degree of correlations. For the 1022 galaxies based on SDSS, the eigenvalues of PC1 through PC6 are 4.52, 0.70, 0.37, 0.20, 0.17 and 0.02 (Figure \ref{pc6}), respectively, where the maximum possible value is 6. Based on common PCA criteria, eigenvalues larger than 1 are considered significant. We therefore only plot PC1 and PC2. Next we conducted PCA without color, and we found the eigenvalues of PC1 through PC5 to be 4.15, 0.41, 0.22, 0.20, and 0.02 (Figure \ref{pc5}), respectively, where the maximum possible value is 5. 


The above observations confirm the finding of D08. All the observed parameters are tightly correlated with PC1. The eigenvalue of PC1 indicates that it can explain 83\% of the variance in the data (when color is not included). Color itself forms a second principal component. This might be explained by the fact that the optical color tends to be strongly affected by recent star formation activities and thus carries extra information that is unrelated to the global formation history of the galaxies. We will come back to the issue of color in Section~4. D08 claimed that the strong dominance of PC1 implies a single physical parameter to govern the structure of galaxies. However, this may be a oversimplified view of galaxies due to limited observational parameters. To further test this, we extend the PCA to the near-infrared.

\begin{figure}[ht]
\begin{center}\includegraphics[width=1\columnwidth]{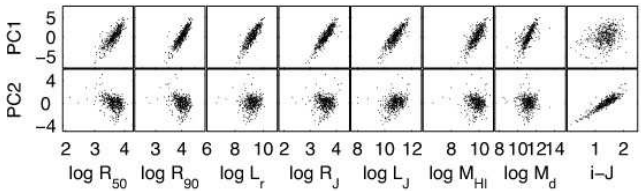}\end{center}
\caption{The PCA results for 481 galaxies from ALFALFA, SDSS, and 2MASS with colors. Here we only show the strongest ones, PC1 and PC2 because other principal components are not significant by PCA criterions. PC1 is well correlated with all the variables. The color is correlated with other variables and PC1 in the first row as well as strongly correlated with PC2 as in Fig \ref{pc6}. In this plot, PC1= 0.37$\log R_{50}$ + 0.39$\log R_{90}$ + 0.39$\log L_r$ + 0.39$\log R_J$ + 0.38$\log L_J$ + 0.36$\log M_{HI}$ + 0.34$\log M_d$ + 0.12($i-J$) and PC2= - 0.33$\log R_{50}$ - 0.25$\log R_{90}$ + 0.08$\log L_r$ -0.01$\log R_J$ + 0.31$\log L_J$ - 0.08$\log M_{HI}$ + 0.01$\log M_d$ + 0.85($i-J$)}. 
\label{pc8}
\end{figure}

\begin{figure}[ht]
\begin{center}\includegraphics[width=1\columnwidth]{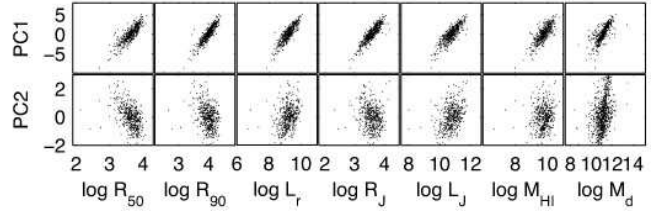}\end{center}
\caption{The PCA results for 481 galaxies from ALFALFA, SDSS, and 2MASS without colors. Here we only show the strongest ones, PC1 and PC2 because other principal components are not significant by PCA criterions. PC1 is well correlated with all the variables, as in Fig \ref{pc5}. In this plot, PC1= 0.38$\log R_{50}$ + 0.40$\log R_{90}$ + 0.49$\log L_r$ + 0.49$\log R_J$ + 0.38$\log L_J$ + 0.36$\log M_{HI}$ + 0.34$\log M_d$ and PC2= $-0.59\log R_{50}$ $-0.39\log R_{90}$ + 0.36$\log L_r$ -0.22$\log R_J$ + 0.46$\log L_J$ + 0.14$\log M_{HI}$ + 0.29$\log M_d$.} 
\label{pc7}
\end{figure}

We included the 2MASS $J$-band radius and the luminosity to investigate the role of stars, which dominate the baryonic matter in galaxies. For the 481 galaxies detected by 2MASS, we found the eigenvalues of PC1 through PC8 to be 5.41, 1.17, 0.52, 0.39, 0.30, 0.16, 0.03 and 0.02 (Figure \ref{pc8}), respectively, where the maximum possible value is 8. Conducting PCA without color again, we found that the eigenvalues of PC1 through PC7 are 5.35, 0.62, 0.45, 0.31, 0.19, 0.05 and 0.02 (Figure \ref{pc7}), respectively, where the maximum possible value is 7. 

The overall trends in the above near-infrared PCA are similar to those in the optical PCA. When color is not included, PC1 dominates and can explain 76\% of the variance in the data. The importance of PC2 slightly increases from 8\% in the optical case to 9\% in the near-infrared case here.  When color is included, it forms another principal component by itself. These again confirm the observations of D08 that only one physical parameter governs the dynamics of galaxies.

A subtle but surprising difference between the optical and near-infrared PCAs is the behavior of color. In the optical PCA in Figure~\ref{pc6}, although the $g-i$ color forms a second principal component, it still weakly correlates with other parameters and it is part of the first principal component. On the other hand, in the near-infrared PCA in Figure~\ref{pc8}, the $i-J$ color almost does not involve in the first principal component and itself forms a second component that is almost independent of other parameters. A potential issue here is the combination of SDSS $i$ and 2MASS $J$. Minor differences between the two photometric systems may hamper the correlations. To test this, we replaced the $i-J$ color with the pure 2MASS $J-K$ color, and we found identical results. In addition, the median errors in $g$ and $i$ for the 1022 SDSS galaxies are 0.0051 and 0.0055, respectively, and the median errors in $i$ and $J$ for the 481 2MASS galaxies are 0.0034 and 0.064, respectively.  These translate to typical color errors of 0.0075 in $g-i$ and 0.064 in $i-J$. Both values are significantly smaller than the color dynamical ranges shown in Figures~\ref{fig_cor6} and \ref{fig_cor8}, meaning that the distribution of the data is not dominated by measurement errors.  We therefore believe that the lack of correlation between the near-infrared color and other parameters is real. We will discuss more on this result in Section~4.

\begin{figure}[ht]
\begin{center}\includegraphics[width=1\columnwidth]{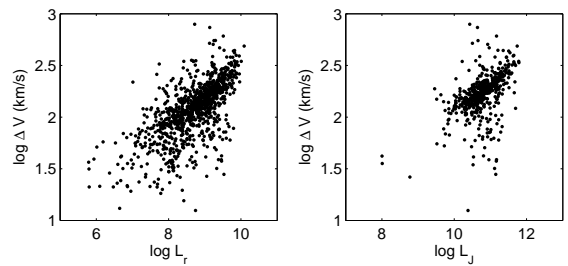}\end{center}
\caption{The Tully-Fisher Relation in $r$ band and $J$ band. The scatter in $J$ band is smaller than that in $r$ band.}
\label{fig_tully}
\end{figure}

Finally, we feel obliged to check the Tully-Fisher relation because the rotational velocity and the luminosity are what we acquired directly. The results are shown in Figure \ref{fig_tully}. We see that the luminosity is proportional to the rotational velocity with the power-law index between 3 and 4 in both $r$ band and $J$ band, which agrees with other results in the literature \citep[e.g.,][]{1992ApJ...387...47P}.
\section{Discussion and Conclusion}
For the 195 galaxies analyzed in D08, the correlations among the parameters $R_{50}$, $R_{90}$, $L_g$, $M_{HI}$ and $M_d$ are obvious. By selecting a much larger, 1022 overlapping samples from SDSS and ALFALFA, we also performed the PCA and confirmed through the high eigenvalues of the correlation matrices that the correlations are similarly strong. It follows that the radius, luminosity, \ion{H}{1} mass and dynamical mass of those chosen galaxies are tightly correlated. However, the color appears to be much less correlated. This indeed complicates the situation and may be explained by a more sophisticated theory that would include, for example, the influence of recent star formation and very luminous young stars. Indeed such studies have already been pursued by many authors \citep[e.g.,][]{1961ApJS....5..233D,1982ApJ...257..527T,2001AJ....122.1861S,2004ApJ...601L..29H,2010ApJ...712.1385C}. 

D08 suggests that the optical color ($g-r$ in their case) consists of two components: the ``systematic'' component that correlates with other parameters (and therefore involves in the first principal component), and the ``rogue'' component that is more or less random (and forms the second principal component). It is tempting to assume that the systematic component comes from the established stellar populations and is related to the global formation history of the galaxies, while the random component is related to the ongoing star formation activities and can be short-lived events in the formation history. Surprisingly, our near-infrared analyses suggest a different story. The color we adopted in our near-infrared PCA is $i-J$. It is less affected by ongoing star formation and should reflect the overall stellar populations in the galaxies (and therefore should better correlate with other global parameters). However, in our near-infrared PCA, this color is even more uncorrelated with other parameters. The reason of this is unclear to us, and we can only conclude that colors in this kind of studies are highly nontrivial.

Putting colors aside, D08 and we found that the basic parameters of galaxies are highly correlated and that there exists only one dominant principal component. This is true both in the optical and near-infrared bands. Based on the hierarchical galaxy formation assumption, the dark matter halo formation has been well studied through the merger tree process \citep[e.g.,][]{1977ApJ...218..333K,1991ApJ...379...52W,longair,2000MNRAS.319..168C,2004ARA&A..42..603K,2006MNRAS.370..645B,2009ApJS..182..216K,2009MNRAS.397.1776F,2010arXiv1001.1484P}. D08 believe that this scenario may not be consistent with the simple relation between all basic galactic parameters because the processes of merger would break the original galactic structures. Although our analysis has confirmed D08's finding, we are not convinced by their interpretation. 

It is well-known that the astrophysics at small, galactic scales involves additional complex physical effects, such as that associated with the non-gravitational baryon physics. Indeed, theories and simulations such as semi-analytical modelings \citep{2005MNRAS.363L..31N,2006RPPh...69.3101B} have shown the important influence of baryon physics in the galactic formation processes. Moreover, remarkable progresses have been made with hydrodynamical simulations in resolving some long-standing issues. For example, the merger driven CDM galaxy formation is known to produce too many elliptical type galaxies. By adding gas components in high resolution hydrodynamic cosmological simulations, \citet{2007MNRAS.374.1479G} showed that disk galaxies can quickly form after major mergers happen. Their simulations also reproduced the Tully-Fisher relation and provided a plausible solution to the missing satellite problem. \citet{2008Science.319.174M} showed that stellar feedback can potentially solve the well know ``cusp'' problem in the dark matter density profile when the simulations have sufficiently high resolutions (see also \citealp{2009ApJ.695.292C,2010Nature.463.203G}). We therefore consider the results of galaxy properties the manifestation of a more sophisticated version of rich baryonic physics under the merger tree scenario.


In summary, recent simulations have shown that baryonic physics can play extremely important roles in determining the structure of galaxies.  It remains to be seen whether the rapid progress in putting baryon in synergy with dark matter can eventually reproduce the simple correlation observed here. We thus find it premature to conclude that the CDM model is necessarily flawed based on the correlation of the global parameters of galaxies found so far.
\acknowledgments
This research is supported by Taiwan National Science Council under Project No. NSC NSC97-2112-M-002-026-MY3 (P. Chen and Y.Y. Chang) and NSC98-2112-M-001-003-MY2 (W.H. Wang and Y.Y. Chang), and by US Department of Energy under Contract No. DE-AC03- 76SF00515. We also thank the support of the National Center for Theoretical Sciences of Taiwan. We are grateful to L.W. Lin and H. Hirashita for interesting and encouraging discussions.
We also thank M. Disney and P. Kroupa for comments that improve the manuscript.

The Arecibo Observatory is part of the National Astronomy and Ionosphere Center, which is operated by Cornell University under a cooperative agreement with the National Science Foundation. 

Funding for the SDSS and SDSS-II has been provided by the Alfred P. Sloan Foundation, the Participating Institutions, the National Science Foundation, the U.S. Department of Energy, the National Aeronautics and Space Administration, the Japanese Monbukagakusho, the Max Planck Society, and the Higher Education Funding Council for England. The SDSS Web Site is http://www.sdss.org/.

The SDSS is managed by the Astrophysical Research Consortium for the Participating Institutions. The Participating Institutions are the American Museum of Natural History, Astrophysical Institute Potsdam, University of Basel, University of Cambridge, Case Western Reserve University, University of Chicago, Drexel University, Fermilab, the Institute for Advanced Study, the Japan Participation Group, Johns Hopkins University, the Joint Institute for Nuclear Astrophysics, the Kavli Institute for Particle Astrophysics and Cosmology, the Korean Scientist Group, the Chinese Academy of Sciences (LAMOST), Los Alamos National Laboratory, the Max-Planck-Institute for Astronomy (MPIA), the Max-Planck-Institute for Astrophysics (MPA), New Mexico State University, Ohio State University, University of Pittsburgh, University of Portsmouth, Princeton University, the United States Naval Observatory, and the University of Washington. 

This publication makes use of data products from the Two Micron All Sky Survey, which is a joint project of the University of Massachusetts and the Infrared Processing and Analysis Center/California Institute of Technology, funded by the National Aeronautics and Space Administration and the National Science Foundation.

\end{document}